# Building a Resilient Cybersecurity Posture: A Framework for Leveraging Prevent, Detect and Respond Functions and Law Enforcement Collaboration.

Francesco Schiliro' [1, 2]

f.schiliro@adfa.edu.au

[1]Australian Defence Force Academy, University of New South Wales, Canberra
[2]Macquarie University

20 March 2023

**Abstract**

This research paper proposes a framework for building a resilient cybersecurity posture that leverages prevent, detect, and respond functions and law enforcement collaboration. The Cybersecurity Resilience and Law Enforcement Collaboration (CyRLEC) Framework is designed to provide a comprehensive and integrated approach to cybersecurity that emphasizes collaboration with law enforcement agencies to mitigate cyber threats. The paper compares and contrasts the CyRLEC Framework with the NIST Cybersecurity Framework and highlights the critical differences between the two frameworks. While the NIST framework focuses on managing cybersecurity risk, the CyRLEC Framework takes a broader view of cybersecurity, including proactive prevention, early detection, rapid response to cyber-attacks, and close collaboration with law enforcement agencies to investigate and prosecute cybercriminals.

The paper also provides a case study of a simulated real-world implementation of the CyRLEC Framework and evaluates its effectiveness in improving an organization's cybersecurity posture. The research findings demonstrate the value of the CyRLEC Framework in enhancing cybersecurity resilience and promoting effective collaboration with law enforcement agencies. Overall, this research paper contributes to the growing knowledge of cybersecurity frameworks and provides practical insights for organizations seeking to improve their cybersecurity posture.

**Key terms:** Cybersecurity, cybercrime, risk management, healthcare systems, information security, I.T. infrastructure, threats, Law Enforcement.

**Introduction**

In the contemporary world of technological evolution, the advent of digital data has led to the emergence of cyberspace threats. These attacks can result in significant financial losses, brand reputation damage, and even sensitive data compromise. Leaders must be at the forefront of implementing proactive prevention and mitigation strategies through effective cybersecurity frameworks. Various cybersecurity frameworks have been developed to guide organizations in building effective cybersecurity programs in response to this growing need.

In this research, the Cybersecurity Resilience and Law Enforcement Collaboration (CyRLEC) Framework will be the main focus of such a security architecture. It is a state-of-the-art model that highlights the role of collaboration in the cybersecurity landscape through prevention, identification, and reaction to threats for improved protection procedures. Thus, it is an influential model regarding applicability to the current digital arena and criminal developments regarding information security.

The purpose of this study is to develop a comprehensive examination of the CyRLEC Framework and its role in fostering strong cybersecurity structures for institutions. The focus will also be on modelling a real-world deployment and assessing its impact on organizational cybersecurity. A comparison to other frameworks in the domain will be highlighted by examining its distinctive aspects.

**Literature Review**

The term "cybersecurity framework" describes an array of policies and procedures organizations adopt to protect their information and mitigate potential threats (Taherdoost, 2022). Implementing a security architecture for an organization requires leaders to conduct an internal and external assessment of its environment. Sometimes, a single standard may not always be enough to fulfill an organization's requirements. As a result, leaders should consider whether or not they need to use various standards. Organizations may thus employ the recommendations in their entirety or part, as needed, or combine them with other protocols to enhance and supplement existing regulations (Taherdoost, 2022). It is also important to note that these frameworks do not represent a universal strategy for controlling cybersecurity incidents for vital infrastructure. Companies will always face distinct risks since they have various threats, weaknesses, and tolerances. They also differ in how they adapt a framework's recommended practices.

In healthcare, cyber-attacks disrupt operations, lead to financial losses, and adversely impact patient care. Most healthcare institutions are not always well equipped to deal with cybercrime incidents, even though these incidents might have severe consequences for patient safety. In the U.K., this is a significant problem affecting most medical facilities. Cybersecurity financing is rarely guaranteed, especially in public sector health systems, making healthcare more susceptible than other vital sectors. Although other industries, such as telecommunication and finance, invest between 4% and 10% of their budgets into I.T. infrastructure, Most NHS Trusts in the U.K. only allocate 1% to 2% (Ghafur, et al., 2019). These statistics highlight the need for the sector to invest in this domain. One such cybersecurity framework is the CyRLEC.

Implementing the proper cybersecurity framework is essential for successfully implementing information security standards. Adopting a security architecture that outlines the scope, application, and assessment procedures and provides a basic structure and technique for securing vital digital assets is one way to ensure adequate cybersecurity. Organizations may benefit from cybersecurity frameworks by using them as a guide for successfully enacting cybersecurity standards and improving their ability to prevent, detect, and react to cyberattacks. Cybersecurity paradigms are adaptable, allowing clients to select what they need from the model, including the methodologies and the technological practices offered. They also provide helpful advice for implementing security protocols inside an enterprise. Flexible cybersecurity architectures allow for lower implementation costs and enable resilience.

Existing information security frameworks and standards, such as NIST, ISO/IEC 27001, and COBIT, are dedicated to addressing cybersecurity issues. ISO/IEC 27001 is a global cybersecurity framework that mandates institutions to develop and enforce thorough and uniform procedures for information security (Culot, Nassimbeni, Podrecca, & Sartor, 2021). These safeguards are meant to reduce potentially harmful outcomes. On its part, NIST details the measures an organization may take to safeguard its resources, from vigilantly monitoring potential risks to implementing swift remedies if an attack has been made (Argaw, et al., 2020). It provides an institutional structure that may be used to evaluate and enhance an establishment's strategy or approach for identifying, protecting, detecting, responding to, and recovering from cybersecurity threats. The primary goal of COBIT is to standardize I.T. governance policies and procedures (Wolden, Valverde, & Talla, 2015). It provides a transparent set of techniques for managing business risks, security processes, and technological challenges, assisting management in its risk management efforts related to I.T. governance (Garba & Bade, 2021). Thus, COBIT is a multifaceted instrument that aids I.T. administrators in coordinating the alignment of business risks, technological challenges, and control requirements.

**The CyRLEC Framework**

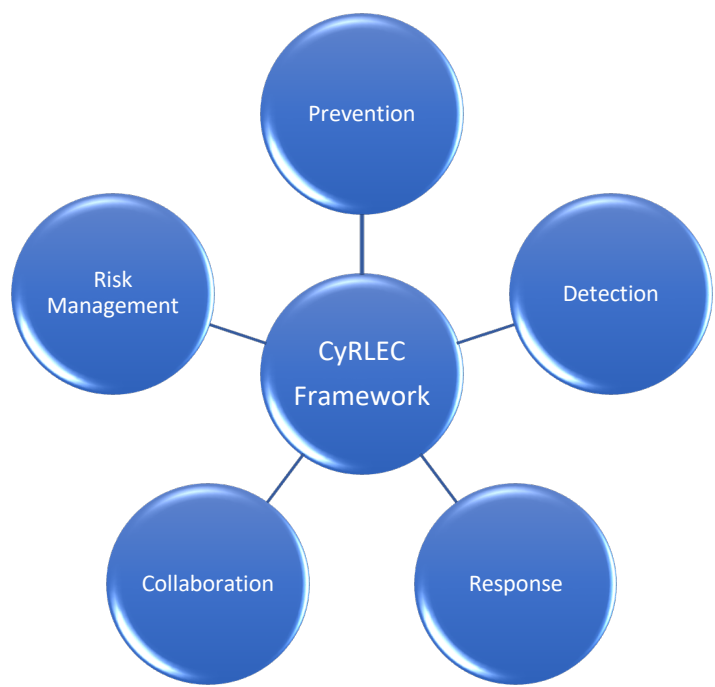

Figure: the five main components of the CyRLEC framework.

The security framework's key elements include risk management, prevention, detection, response, and collaboration. Together, these components promote high levels of cybersecurity in organizations. The model comprises five main components: risk management, prevention, detection, risk response, and collaboration with various institutions. Some of its tenets are similar to those enforced in NIST standards. However, NIST lacks the collaborative aspect of the CyRLEC framework. The latter stresses the value of cooperation between businesses and authorities to strengthen cybersecurity. It involves exchanging information about potential threats, coordinating incident response, and creating joint training and awareness initiatives. It also emphasizes the value of developing strong relationships with law enforcement organizations, including configuring lines of contact and participating in regional cybersecurity steering committees.

The overarching goal of the CyRLEC Architecture is to equip institutions with robust cybersecurity architecture in a structured and feasible manner. Enterprises can strengthen their defences against cyber-attacks and reduce the fallout from an incident by using prevention, detection, and response measures and working with law enforcement. As the world advances in technology, the digital landscape evolves, giving rise to new and more sophisticated threats. Therefore, adaptable, and comprehensive security models are crucial in this environment. CyRLEC highlights the role of law enforcement agencies in the cybersecurity spectrum through reporting (Curtis & Oxburgh, 2022). Law enforcement organizations are familiar with the strategies, methods, and practices malicious attackers use. Additionally, they have access to information and resources that give them a great deal of understanding about certain crimes. Since cybercrime networks commonly operate internationally, law enforcement should be involved in securing data. CyRLEC emphasizes the need for organizations to foster a culture of cooperation with these agencies by exchanging vital information. Law enforcement agencies and the healthcare sector must work together to respond to organized cybercrime. Thus, by taking a proactive approach, the private sector can help law enforcement play a crucial role in anticipating and preventing cybercrimes rather than merely reacting (Curtis & Oxburgh, 2022).

The healthcare sector is one of the most lucrative targets for cybercriminals (Jalali & Kaiser, 2018). Hackers can access and steal patient information, including their names, financial details, and medical records (Argaw, et al., 2020). Facilities across the industry are in a precarious position due to the prevalence of cyberattacks and the historically low level of security spending in the sector. Leaders in the industry need to implement robust information security procedures to keep their data safe, and using a trusted security framework like CyRLEC can simplify the process. The proposed security architecture offers a comprehensive approach to ensuring the security of information systems through proactive strategies.

**Methodology**

The CyRLEC framework is applicable to various industries, including healthcare. The current study uses a simulated model of real-world deployment in hospitals, where cybersecurity and data protection are significant concerns. The study utilizes a qualitative design with semi-structured interviews, and eligible participants include personnel with the necessary skills and experience to manage the institution's information security, including patient data and electronic health records. The purpose is to obtain feedback from the facility's leadership, including the CEO, chief administrative and finance officers, chief technology officer, I.T. security officer, data security specialists, and hospital staff dealing with patient data, regarding cybersecurity. The participants are contacted via email and provided with an informed consent form outlining the study's objectives, purpose, data collection procedure, participant rights, benefits, and risks if any. Confidentiality is crucial, and the responses are coded anonymously (Rodriguez, et al., 2022).

The interview questions are based on the employees' knowledge of cybersecurity practices and data protection within the hospital and their roles in interacting with and managing patient data. The authors reduced confirmation and interviewer bias by adhering to the guidelines for conducting semi-structured interviews (Bergelson, Tracy, & Takacs, 2022). Thematic analysis was used as the primary data analysis tool, and inductive coding procedures were employed to generate recurring themes relevant to the healthcare organization's cybersecurity framework (Ramanadhan, Revette, Lee, & Aveling, 2021). Coding helped the authors categorize the data into meaningful sets to identify the key variables, including risk management procedures, prevention, detection and response strategies, existing infrastructure, and employee training. Participation from various departments ensured that the information represented different perspectives and dimensions of data security across the facility.

**Results**

The findings from the simulation can be applied to other sectors as well. The study identified several emerging themes, including risk management, collaboration, and prevention, detection, and response systems. The results suggest that a practical framework can help ensure that all relevant parties, processes, and technology are in place to deal with information and cyberspace threats in a manner that aligns with the hospital's goals and adheres to best practices, regulatory, and legislative standards. CyRLEC can aid healthcare institutions in preventing, transferring, mitigating, or accepting risks associated with their use of information. Additionally, it ensures that all aspects of the security program, including policy design, execution, assessment, surveillance, and reporting processes, are adequately addressed. The framework provides enterprises with the necessary structure and flexibility to manage cybersecurity risks strategically. Healthcare institutions face different levels of data breaches, exposures, and tolerance. They can adopt the concepts, best practices, procedures, and recommendations offered in the CyRLEC based on their specific needs. Furthermore, the framework enables hospitals to combat cyber risks more effectively by collaborating with other private and public sector entities, such as law enforcement agencies.

**Discussion**

One of the primary strengths of the CyRLEC framework is its ability to provide a comprehensive and adaptable system that ensures long-term risk management and information security across various organizations, including healthcare. CyRLEC adopts a risk-based operational approach to identify threats and vulnerabilities and develop a resilient information security model that can prevent attacks on an organization's information infrastructure. The ongoing process helps identify new and emerging issues and mitigates them through a comprehensive risk management plan, which includes periodic vulnerability assessments to identify potential weaknesses and loopholes.

As new healthcare solutions and technologies emerge, the need for effective risk management in cybersecurity becomes increasingly pressing. CyRLEC is centred on the risk management field, which encompasses all activities related to detecting, evaluating, and preventing network security threats. Vigilance is required in these operations to monitor the threat environment, identify perceived risks, rank the relative seriousness of dangers, and learn from past occurrences. CyRLEC helps healthcare organizations design and manage risk-reducing systems and procedures to ensure they are prepared to tackle any cybersecurity threat.

Compared to other frameworks like NIST, CyRLEC has a unique advantage in leveraging cooperation with law enforcement agencies. Through collaboration, an institution can gain critical, non-public security information that may help CyRLEC understand the weaknesses leveraged in a security breach, its source, and its intended motive. These benefits of collaboration have been demonstrated across various industries over the years. For example, in partnership with the Justice Department, the FBI recovered $500,000 in ransomware gains from North Korean cybercriminals who had attacked healthcare facilities across the U.S. (The United States Department of Justice, 2020). The state-sponsored cell spreading ransomware was disrupted thanks to swift reporting and collaboration. This case illustrates how CyRLEC is taking a collaborative and comprehensive approach to combating cybercrime by working closely with law enforcement agencies.

However, a significant limitation of the CyRLEC framework is its failure to account for the constantly evolving cybercrime space. The lack of training is a critical issue affecting this area (Chowdhury & Gkioulos, 2021). Hackers' strategies, tactics, and processes continue to evolve (He, Aliyu, Evans, & Luo, 2021), and therefore the framework needs to be continuously updated and patches provided to seal emerging loopholes in the system. Without continuous learning, the current framework may not effectively mitigate future attacks. Additionally, adopting and implementing the framework can be costly for institutions. Since the benefits of its implementation may not be immediately tangible and the costs may exceed the returns, organizations may not readily embrace it. Moreover, the architecture is complex to set up, requiring the purchase of infrastructure and the hiring of security specialists to manage it.

**Conclusion**

The rapid and widespread adoption of technology has increased susceptibility to cyber risks, compromising patient safety and the security and privacy of health data. One effective way to mitigate these risks is by implementing the CyRLEC framework, which is designed around prevention, detection, and response models. One of the main benefits of CyRLEC over other frameworks is its ability to coordinate with law enforcement, providing an enhanced layer of protection for organizations. Research indicates that CyRLEC is an effective architecture for managing and controlling hospital risks by partnering with policing agencies to address information security risks. Organizations looking to improve their architecture should allocate resources to invest in cybersecurity frameworks for effective and up-to-date risk management.